\documentclass[11pt]{article}   	% use "amsart" instead of "article" for AMSLaTeX format
\usepackage{geometry}                		% See geometry.pdf to learn the layout options. There are lots.
\usepackage{amsmath, amssymb}
\usepackage[round,authoryear]{natbib}
\usepackage{graphicx}
\usepackage[colorlinks=true,linkcolor=blue,citecolor=blue,urlcolor=blue]{hyperref}
\usepackage{tabularx,booktabs,ragged2e,makecell,caption,placeins,float}

\newcolumntype{Y}{>{\RaggedRight\arraybackslash}X} % ragged-right X
\usepackage{placeins}

\hypersetup{
  unicode=true,
  pdftitle={A Taxonomy of Hierarchical Multi-Agent Systems: Design Patterns, Coordination Mechanisms, and Industrial Applications},
  pdfauthor={David J. Moore},
  pdfsubject={Hierarchical multi-agent systems taxonomy},
  pdfkeywords={multi-agent systems, hierarchy, coordination, auction, consensus, contract net, industry}
}

\usepackage{bookmark}

\title{A Taxonomy of Hierarchical Multi-Agent Systems: Design Patterns, Coordination Mechanisms, and Industrial Applications}
\author{David J. Moore \\
\small Independent Researcher, Houston, TX, USA \\
\small \texttt{david@davidjmoore.com}
}

\date{\today}

\begin{document}

\maketitle

\begin{abstract}
Hierarchical multi-agent systems (HMAS) organize collections of agents into layered structures that help manage complexity and scale. These hierarchies can simplify coordination, but they also can introduce trade-offs that are not always obvious. This paper proposes a multi-dimensional taxonomy for HMAS along five axes: control hierarchy, information flow, role and task delegation, temporal layering, and communication structure. The intent is not to prescribe a single “best” design but to provide a lens for comparing different approaches.

Rather than treating these dimensions in isolation, the taxonomy is connected to concrete coordination mechanisms—from the long-standing contract-net protocol for task allocation to more recent work in hierarchical reinforcement learning. Industrial contexts illustrate the framework, including power grids and oilfield operations, where agents at production, maintenance, and supply levels coordinate to diagnose well issues or balance energy demand. These cases suggest that hierarchical structures may achieve global efficiency while preserving local autonomy, though the balance is delicate.

The paper closes by identifying open challenges: making hierarchical decisions explainable to human operators, scaling to very large agent populations, and assessing whether learning-based agents such as large language models can be safely integrated into layered frameworks. This paper presents what appears to be the first taxonomy that unifies structural, temporal, and communication dimensions of hierarchical MAS into a single design framework, bridging classical coordination mechanisms with modern reinforcement learning and large language model agents.

\end{abstract}

\noindent \textbf{Keywords:} hierarchical multi-agent systems, coordination mechanisms, taxonomy, reinforcement learning, industrial AI, smart grids, oil and gas, human-AI collaboration

\section{Introduction}
Modern applications of artificial intelligence often involve multiple autonomous agents that must work together in a complex environment. Hierarchical multi--agent systems (HMAS) have emerged as an effective architectural paradigm to coordinate large collections of agents by organizing them into layers or organizational structures. In an HMAS, higher--level agents (or agent teams) typically oversee or coordinate lower--level agents, creating a chain of command analogous to human organizational hierarchies. This layered approach is motivated by several factors. First, hierarchy addresses scalability: as the number of agents grows, a purely flat (fully decentralized) organization may struggle with communication overhead or global coherence. By delegating decision--making to intermediate “leader” agents, the system can manage complexity through divide--and--conquer strategies. Hierarchical mechanisms assign greater responsibilities to some agents, allowing flexible adaptation to task demands and efficient management of large--scale MAS. In autonomous driving, for example, researchers have proposed selecting regional leader agents that coordinate local traffic while a top--level agent optimizes network--wide objectives. Second, hierarchy enables different levels of abstraction and temporal scales in decision--making. High--level agents can plan in broader time horizons or abstract state spaces (e.g. mission planning), while lower--level agents execute detailed actions (e.g. motion control), improving overall coherence. Third, structured hierarchies can facilitate more organized coordination and conflict resolution. They establish clear authority relationships (who guides whom) and defined communication channels, reducing indecision or oscillations that might occur in fully egalitarian agent teams. As Malone and Crowston’s classic definition suggests, coordination is fundamentally about managing interdependencies between activities – a task made easier when there is some form of structured oversight or leadership \citep{Malone1994multi}.

Early work in distributed AI and multi--agent systems recognized the impact of organizational design on performance. Organizational paradigms such as hierarchies, teams, coalitions, holarchies, and markets were studied to understand their trade--offs. Notably, hierarchical organizations (often implemented via manager–worker patterns or tree topologies) were shown to improve global efficiency at the cost of some robustness, whereas fully decentralized “team” organizations maximize resilience and equality but can be less efficient in large groups. Recent surveys highlight a renewed interest in hybrid approaches that combine hierarchical and decentralized coordination to get the best of both worlds. Recent industry investments demonstrate an increasing interest: multi-agent systems garnered \$12.2 billion in funding through more than 1,100 transactions in Q1 2024, signifying sustained confidence in the transformative potential of MAS across billion-dollar sectors such as healthcare, mobility, finance, and defense \citep{aalpha2024}. A thorough 2025 study by \cite{Sun2025multi} finds ``hybridization of hierarchical and decentralized mechanism'' as a crucial strategy for achieving scalability while maintaining adaptability, hence validating the significance of this papers categorization approach. Indeed, hierarchical structures are ubiquitous in nature (e.g. social insect colonies with worker castes and queens) and human society (organizational charts in enterprises), suggesting that well--designed hierarchies can greatly enhance coordination in multi--agent AI systems.

In addition to structural benefits, HMAS also align well with how humans interact with AI systems. In complex mission--critical domains like energy operations or military command centers, it is natural to have a hierarchy where human supervisors oversee AI agents, which in turn manage lower--level automated processes. This not only mirrors real management structures but also provides a point at which human expertise and intent can be injected into the loop. Hierarchies can thus facilitate human--agent collaboration by defining clear roles for human operators (e.g. strategic decision--maker) and autonomous agents (e.g. tactical executors). However, such integration also raises challenges – the system must provide explainability upwards (so that humans understand the agents’ decisions) and trust mechanisms to ensure humans confidently delegate to agents .

Given the growing scale and complexity of multi--agent applications – from smart grids with hundreds of distributed energy resources, to autonomous vehicle fleets, to adaptive supply chains – there is a need for a systematic taxonomy and design guidance for hierarchical multi--agent systems. While previous taxonomies of multi--agent systems exist (e.g. classifying by team size, communication topology, agent heterogeneity), this work proposes an updated taxonomy focused specifically on hierarchical organization patterns and their coordination mechanisms. This paper builds on insights from classical frameworks (like the Contract Net Protocol by \cite{smith1980} for task allocation and organizational theory in MAS) as well as recent advances (such as hierarchical multi--agent reinforcement learning and large language model--based agents). The goal is to outline the design space of HMAS in a multi--dimensional way, clarifying the choices and trade--offs for system architects. The taxonomy is grounded in real--world applications, particularly in the energy and operations domains where the author has deep expertise. By tying each taxonomy dimension to concrete industrial case studies (e.g. hierarchical MAS for oil well monitoring, or for microgrid control), it demonstrates the practical relevance of these design patterns. Finally, open research issues are identified that emerge at the intersection of hierarchical structures with modern AI capabilities – for example, how to maintain global transparency in a deeply layered agent system, or how to incorporate the reasoning power of LLM--based agents into traditional hierarchies. This work contributes a unifying taxonomy that spans five dimensions — control, information flow, role/task delegation, temporal layering, and communication structure and explicitly connects them to coordination mechanisms and industrial deployments. By doing so, it advances beyond prior surveys to provide a practical design lens for both researchers and system architects

\section{Taxonomy of Hierarchical Multi--Agent Systems}

To characterize the design space of hierarchical MAS, this paper introduces a taxonomy comprising five key axes. Each axis represents a fundamental aspect of how a multi--agent hierarchy can be structured or how it operates. Each dimension of classification is examined in terms of its spectrum of possibilities (often with two or more extremes), with representative systems or methods illustrated and the associated advantages and trade--offs emphasized. These axes are not entirely independent – in fact, certain design patterns (discussed along the way) involve particular combinations of these attributes. Nonetheless, isolating the axes allows for greater clarity about the unique contribution of each aspect to the overall system design.

The five axes are: (1) Control Hierarchy, (2) Information Flow, (3) Role and Task Delegation, (4) Temporal Hierarchy, and (5) Communication Structure. Each is summarized below.

\subsection{Control Hierarchy: Centralized vs. Decentralized vs. Hybrid}
\label{subsec:control hierarchy}

One defining feature of an HMAS is the pattern of control authority – i.e. how decision--making power is distributed among agents. On one end of this spectrum, a fully centralized hierarchy has a single top--level agent (or a small set of high--level agents) that make most decisions and directly instruct lower--level agents. Traditional planning systems often adopt this model: a central “brain” computes an assignment of tasks or resources and sends commands to many executor agents. This can be efficient for global optimization but suffers from a single point of failure and limited robustness. At the opposite end, a fully decentralized system has no single leader; all agents are more or less equal, and any hierarchy is only implicit (perhaps arising from dynamic interactions). Decentralized multi--agent systems rely on mechanisms like consensus, local voting, or emergent alignment of agent behaviors to achieve coordination without a central authority. Such systems are robust to failures and adaptable, but they may have difficulty achieving coherent global behavior as the group size grows.

In practice, many multi--agent systems blend these extremes into hybrid control hierarchies. A hybrid HMAS might have several layers: e.g., a top layer of coordinators that set high--level goals, and a bottom layer of workers that carry out tasks, with possibly a middle layer aggregating information. Hybrid architectures aim to reap the benefits of both approaches – scalability and global oversight from hierarchy, plus resilience and responsiveness from decentralization. Indeed, recent research explicitly encourages marrying hierarchical and decentralized coordination as a way to handle large, dynamic environments. The intuition is that some structure (even a partial hierarchy) can guide the swarm of agents more effectively (preventing deadlocks or global inefficiencies), while decentralized elements ensure that if one leader fails, the system can still continue and agents can adapt locally.

Design patterns: A classic pattern exemplifying centralized control is the Contract Net Protocol (CNP) \citep{smith1980} . In CNP, a distinguished manager agent announces tasks and collects bids from other agents (contractors), then assigns the task to the best bidder. This reflects a star--like hierarchy: the manager has the authority to delegate tasks, though the process of bidding adds flexibility. Conversely, a pattern for decentralized control is the consensus algorithm, where each agent adjusts its state (e.g., heading direction, or agreement on a value) based on neighbors’ states until eventually all converge. There is no fixed leader in consensus – any agent can influence the outcome, and leadership may effectively emerge if some agent has a critical piece of information (e.g., an informed agent in flocking or a high--confidence estimate in sensor fusion). Hybrid patterns often involve leader election or dynamic authority: for instance, agents might vote or use heuristics to select a leader for the current situation, who then acts in a centralized manner until the task is done. In autonomous vehicle networks, a temporary leader might be chosen to coordinate a platoon of cars, after which control reverts to individual vehicles. Another example is hierarchical reinforcement learning in multi--agent teams: \cite{Ahilan2019multi} Feudal Multi--Agent Hierarchies (FMH) framework explicitly trains a high--level manager agent to communicate goals to multiple low--level worker agents. The manager in FMH maximizes the overall reward and sets sub--goals, while workers are rewarded for achieving those sub--goals. This approach improved performance and scalability compared to fully centralized or fully distributed learning, demonstrating the power of a judicious control hierarchy in cooperative multi--agent reinforcement learning.

Recent advancements in hierarchical consensus systems further validate this strategy. To address the limitations of the Centralized Training with Decentralized Execution (CTDE) paradigm, \cite{feng2024} introduced the Hierarchical Consensus-based Multi-Agent Reinforcement Learning (HC-MARL) framework, which use contrastive learning to foster global consensus across agents. Utilizing consensus as ancillary information to guide collaborative activities during execution enables cooperation without direct communication. 

In summary, the control hierarchy axis ranges from strongly centralized (one or few leaders directing all) to purely decentralized (no fixed leaders), with many gradations in between. Choosing the right balance is critical: centralized control simplifies global coordination but can bottleneck, whereas decentralization enhances robustness and parallelism at the cost of potentially slower convergence to global optima. Many successful HMAS implementations opt for a middle ground – e.g., a multi--tier architecture where strategic decisions are made at the top and tactical decisions are decentralized at the bottom. This pattern reappears in industrial case studies (Section \ref{sec:HMAS in Industry}), such as power grid management where regional controllers handle local optimizations under the guidance of a central dispatch center.

\subsection{Information Flow: Top--Down, Bottom--Up, and Peer--to--Peer}

Closely related to control hierarchy is the information flow within the agent organization. This axis concerns how knowledge, data, and directives circulate in the system. In a top--down flow, information primarily moves from higher layers to lower layers – for example, a global planner agent might compute a plan and then disseminate specific instructions to each subordinate agent. This mode aligns with centralized control and is efficient for broadcasting goals or commands. Bottom--up flow, on the other hand, means that information chiefly rises from lower--level agents up to the higher levels. For instance, local agents might observe their environment or partial results and report summaries or alerts upward to a supervisor agent that aggregates these into a global picture. Many monitoring and diagnosis systems use bottom--up information flow: sensors (agents) feed data to an aggregator or decision module. Finally, peer--to--peer flow refers to information being shared laterally among agents at the same layer, without necessarily funneling through a hierarchy. Peer communication is prevalent in swarm systems and distributed consensus approaches, where agents continuously exchange state with neighbors.

Real hierarchical MAS often employ a mix of flows. Consider a real--time operations center scenario: field agents (low level) send status reports upstream (bottom--up) to regional coordinators, who then issue guidance or adjustments downstream (top--down) to the field agents. Additionally, field agents might also communicate among themselves for local coordination (peer--to--peer) if allowed. The design of information flow has a large impact on performance. Top--down flow ensures directives are consistent and agents know the global mission, but if lower agents don’t communicate upward, the leaders may become blind to ground realities. Bottom--up flow improves situational awareness at higher layers, enabling adaptive top--level decisions, but too much upward data can overwhelm the leader (creating an information bottleneck). Peer--to--peer flow can solve local problems without troubling higher--ups, but excessive lateral chatter might cause confusion or require complex consensus protocols.

Design patterns: A quintessential top--down pattern is the blackboard system – here a central data store or “blackboard” is updated by higher--level reasoning modules, and lower--level agents simply follow the posted plans or tasks. The flow of final instructions is downward. In contrast, a distributed sensing network operates bottom--up: numerous simple agents (sensors) push data up to a fusion center or collectively compute an aggregate (e.g. averaging). Publish--subscribe architectures allow a blend: an agent can publish information that either flows up to any subscriber above, or horizontally to peers subscribed to that topic. Hierarchical planning systems (like layered architectures in robotics) explicitly design information flows: the deliberative layer sends goals to the executive layer, which in turn sends commands to the reactive layer. Feedback flows upward as each layer reports status or failure back to its superior. This is a classic top--down with bottom--up feedback loop. On the other hand, market--based coordination in MAS (such as auctions for tasks) often uses peer--to--peer information exchange: agents communicate bids and offers among themselves (or to a mediator) rather than rigidly upward or downward.

An HMAS can also be characterized by whether its communication network is explicitly hierarchical (e.g., tree topology for message passing) or whether hierarchy is only in control authority but not in info routing. Some multi--agent frameworks impose a tree network (each agent only talks to its parent or children), inherently enforcing top--down and bottom--up flows. Others allow free communication among certain roles. For example, in a holonic MAS (nested hierarchy), lower--level “holons” encapsulate information locally but can share summaries with a higher holon; meanwhile, holons at the same level might coordinate peer--to--peer if they belong to the same parent. Thus, designing information flow goes hand--in--hand with defining the interfaces between layers in an HMAS.

In this taxonomy, information flow is treated as an independent axis to emphasize that even a strongly centralized control system might still benefit from bottom--up sensing, and even a decentralized control system might implement some top--down broadcasting (e.g., a common world state shared to all agents). Effective HMAS leverage multiple flows: top--down communication for spreading global intents, bottom--up communication for aggregating distributed knowledge, and lateral communication for efficient local interactions. Balancing these flows is crucial for maintaining both global awareness and local reactivity. As noted in organizational theory, an agent organization influences data flow and coordination patterns – a rigid hierarchy might restrict peer communication (preventing redundancy but also potentially hindering cooperation), whereas a lax structure might produce information overload. The taxonomy encourages designers to ask: how will information propagate in my agent hierarchy, and does that support the desired coordination mechanism? Section 3 discusses coordination algorithms, illustrating how specific algorithms presuppose particular information flow patterns. For example, distributed consensus relies on peer-to-peer message passing, whereas task allocation may assume top-down announcements.

\subsection{Role and Task Delegation: Fixed vs. Emergent Roles}

Another important dimension is how roles and tasks are distributed among agents, and whether these allocations are predetermined or can change dynamically. In many hierarchical systems, agents are designed with specific roles in mind – e.g., in a rescue scenario, one agent is the team leader, some are explorers, others are medics. These roles often correspond to differing capabilities or responsibilities (the medic agent handles injured civilians, the explorer drones scout areas, etc.). Fixed--role hierarchy means that the role of each agent (and hence its place in the hierarchy) is defined a priori and remains relatively static. For example, in a refinery operations MAS, you might permanently designate one agent as the refinery coordinator and others as unit controllers or sensor agents. Fixed roles simplify design and can embed domain knowledge (as each role can be programmed/tuned for its function), but they reduce flexibility – the system might not adapt well if an agent fails or if the task distribution shifts.

In contrast, some HMAS employ emergent or dynamic roles, where an agent’s role is not hard--coded but instead emerges through learning or negotiation. In such systems, all agents might start homogeneous, but during operation they differentiate – perhaps one agent takes on a leadership role because it has the most information or the best vantage point, and later hands off that role when conditions change. Recent advances in multi--agent reinforcement learning have showcased this concept. \cite{Wang2020multi} introduce ROMA (Role--Oriented MARL), a framework in which roles are not predefined by the designer but are learned as latent embeddings. Agents with similar learned roles exhibit similar behavior and specialize in certain sub--tasks, and the roles can adapt on the fly. This dynamic role assignment improved performance on complex team tasks by allowing the agents to coordinate implicitly – essentially discovering “who should do what” in an optimal way. Similarly, other studies on emergent roles in robot teams (e.g., in RoboCup soccer, where agents can swap offensive/defensive roles as needed) demonstrate that role flexibility can increase a team’s robustness and efficiency in unpredictable environments.

In more classic terms, the question is whether the task delegation in the hierarchy is explicitly engineered or can self--organize. Fixed delegation fits a top--down command style: tasks are assigned by leaders to specific subordinates known to fill that role. For instance, a factory MAS might always delegate maintenance tasks to the “maintenance agent” and scheduling tasks to the “scheduler agent.” On the other hand, in an emergent delegation scenario, agents might bid or volunteer for tasks regardless of a fixed role. One example is a dynamic task allocation scheme where any agent can take leadership for a task if it currently has capacity, or any drone can decide to become the “base station” if it senses that communication is needed. These emergent behaviors often rely on underlying mechanisms such as distributed auctions, teamwork models, or adaptive learning.

Design patterns: The manager–worker pattern is typically implemented with fixed roles: a manager agent delegates subtasks to worker agents who execute them. Many industrial agent systems follow this, mapping managerial roles to human job roles (e.g., a “foreman agent” vs “worker agents”). In contrast, an emergent leader election pattern would allow any agent to become the manager depending on circumstances (for example, the first robot that finds a target could become the coordinator for retrieval). Contract net (mentioned earlier) is somewhere in between – the initiator of a contract plays a temporary manager role for that task, but which agent is initiator can vary by task, so roles are fluid on a per--task basis. Role dynamics can also be seen in foraging swarm examples: if one robot in a swarm finds food, it might take on the role of recruiter (bringing others) which was not a permanent designation.

From a software engineering perspective, methodologies like GAIA \cite{wooldridge2000gaia, zambonelli2003gaia} or AGR (Agent--Group--Role modeling) \cite{zambonelli2003gaia} introduced explicit role abstractions in MAS design, expecting developers to specify roles and their interactions at design time. This falls in the fixed role camp. However, adaptive MAS research argues for run--time role reallocation – agents might shift roles to balance load or cover for failures. The complexity of designing such systems is higher, but machine learning is providing tools to automate role assignment. As noted by Wang et al., role--based approaches historically required significant domain knowledge and hand--coded role definitions, whereas learning--based approaches provide more flexibility at the cost of needing more experience data. In ROMA, they combine both: they allow roles to emerge (learned), meaning agents with similar roles naturally converge to similar behavior, achieving the benefits of role specialization without rigid prior definitions.

In the proposed taxonomy, Role/Task Delegation is treated as an axis: one end is static role hierarchy (predefined organizational chart, each agent has a fixed position and task scope) and the other end is dynamic role assignment (the hierarchy can reshuffle, or roles appear/disappear as needed). Both extremes have merits. Static roles often make verification and explainability easier – stakeholders know “Agent A is responsible for X” – which is important in safety--critical fields. Dynamic roles promise greater adaptivity and often better optimality in task allocation since the system can reconfigure on the fly. However, dynamic role systems need robust mechanisms to prevent chaos (who decides roles? how to avoid oscillations?), and to maintain identifiability (the system’s behavior should not become unpredictable to human overseers simply because roles keep changing). Recent research on explainable MARL and trust in MAS touches on this: if agents change roles, they may need to explain their new behavior to humans for trust.

\subsection{Temporal Hierarchy: Long--Horizon vs. Short--Horizon Decision Layers}

Hierarchies can also be distinguished by the timescales at which different layers operate. In many complex systems, there is a natural stratification of decision--making by temporal horizon: high--level decisions cover a longer time span or larger anticipation window, while low--level decisions are myopic and quick. The Temporal Hierarchy axis is defined to capture this separation between long--term planning and short--term reaction.

In a temporal hierarchical MAS, an upper--layer agent might set goals that are meant to be achieved over an extended period (hours, days, or an entire mission), whereas lower--layer agents continuously make short--term decisions (seconds or milliseconds) to execute and adjust actions toward those goals. For example, in a hierarchical drone swarm, a top--layer agent could assign each drone a sector of a search area (a decision affecting the next 30 minutes of operation), while the individual drone agents make second--by--second path planning choices to systematically cover their sectors and avoid obstacles. Similarly, in a manufacturing MAS, a high--level scheduler agent plans the production schedule for the day (long horizon), and machine agents control minute--by--minute operations on the factory floor (short horizon), possibly adjusting in real--time to machine faults or quality issues.

Temporal layering is often associated with the concept of hierarchical planning and control in robotics and AI. Techniques like Hierarchical Task Networks (HTN) explicitly decompose a complex task into subtasks recursively, effectively creating a plan tree where high--level tasks (long duration) break down into short--duration actions. In multi--agent settings, one can assign different agents to different layers of such a decomposition. A concrete illustration is the use of macro--actions or options in multi--agent reinforcement learning: some agents (or agent modules) choose a macro--action that may last for several time steps (long horizon decision), while other agents execute primitive actions every time step. \cite{handler2023} proposed a framework where each agent communicates a long--term intention that remains fixed for a while, instead of constantly communicating every tiny action. In their hierarchical cooperation framework, an agent’s high--level policy outputs a latent embedding (a representation of its long--term goal or plan), which is shared with other agents and held constant until the agent changes macro--action. Low--level behaviors are then guided by both this internal long--term intention and short--term observations. Such designs significantly reduce communication needs (since agents only share their big picture intent occasionally, rather than every step) and improve coordination, because each agent can predict teammates’ near--future behavior from their stated long--term intentions. This exemplifies how separating timescales – long--term info for alignment, short--term for execution – can lead to more coherent multi--agent teamwork.

Design patterns: The sense–plan–act hierarchy in classic robotics is a temporal hierarchy: sensing and acting occur in tight loops (fast, real--time), while planning is a slower loop that provides guidance to the lower loops. Many three--layer architectures (e.g., 3T, ATLANTIS) explicitly delineate a deliberative layer (slow, strategic), an executive layer (mid--level, tactical), and a reactive layer (fast, reactive control). When these layers are implemented as separate agents or processes, this constitutes a temporal multi--agent hierarchy. Each layer--agent deals with different time granularity. Another pattern is the monitor–control hierarchy seen in process control systems: a monitoring agent might watch for trends or faults over time (needing to integrate data over long windows), and when it detects something, it triggers a control action in a lower--level agent that adjusts a device immediately.

In multi--agent games or operations research problems, one finds analogous stratifications: e.g., in a multi--agent wargame simulation, a commander agent plans the overall battle (long--term strategy), while individual unit agents make instantaneous combat decisions (short--term tactics). The commander might only update its strategy occasionally (or when major events occur), whereas units act continuously. This improves efficiency by not burdening the commander agent with reacting to every minor fluctuation – that’s delegated to the unit agents’ short--term policies.

From a taxonomy viewpoint, Temporal Hierarchy ranges from systems where all agents operate on the same timescale (no temporal layering) to systems where there are clearly distinct layers each with their own update frequency or planning horizon. For instance, a swarm of identical robots all making decisions at 1 Hz with no higher--level planner has no temporal stratification. Meanwhile, a hierarchical MAS for supply chain management might have a high--level planning agent optimizing weekly logistics, mid--level agents handling daily scheduling, and low--level agents managing real--time warehouse operations – a deeply layered temporal hierarchy.

Implications: Temporal hierarchies can dramatically improve coordination by allowing high--level agents to abstract away the micro--details and focus on long--term dependencies, while low--level agents ensure responsiveness to immediate conditions. However, they introduce challenges in alignment – the higher--level plan must be translated properly into low--level actions, and conversely, low--level events (like emergencies) must propagate up in a timely manner to possibly adjust the high--level plan. This relates back to the information flow axis (Section 2.2): effective temporal hierarchies need good bottom--up signaling (so that long--term plans remain grounded in reality). Additionally, learning in a hierarchical temporal system (like multi--level RL) is complex; one needs techniques to train agents that operate at different time scales (e.g., using reward decomposition or hierarchical credit assignment).

A notable recent approach in MARL is the use of a manager–worker with time model: Ahilan \& Dayan’s FMH (mentioned earlier) had a manager agent set subgoals at a slower pace and worker agents execute primitive actions every step. They found that this approach not only yields better performance but also scales to more agents because the manager can coordinate many workers by giving each a coarse objective, instead of micromanaging each step. The separation of temporal scales meant the manager’s action space (choice of subgoal) was much smaller and easier to optimize over a long horizon, while the workers handled the finer details.

In summary, temporal hierarchy is a powerful axis in designing HMAS. Practically every complex multi--agent application with a notion of planning can benefit from layering by timescale. When mapping out a solution, designers should ask: Can we split the decision process into slower strategic decisions and faster reactive decisions? If yes, assigning these to different agents or agent layers can simplify each agent’s job and improve collective performance. Later sections on applications will note where such temporal layering is used (for example, Section 4.1 on smart grids will discuss long--term scheduling vs. real--time balancing in an MAS for energy management).

\subsection{Communication Structure: Static vs. Dynamic Networks}

The final axis in our taxonomy deals with the structure of communication among agents – specifically, whether the network of who can talk to whom is fixed or can change over time. In any multi--agent system, agents interact over some communication graph (which could be physical, like wireless connectivity, or logical, like subscription to messages). In a static communication structure, the links between agents are predetermined and remain constant (or mostly constant). For example, in a two--level hierarchy you might have a fixed tree: each low--level agent is permanently linked to one high--level agent (its supervisor), and supervisors are linked to a central coordinator. Many organizational MAS assume this kind of static topology, often reflecting an org chart or a network layout that doesn’t change. Static structures are easier to manage and analyze; one can guarantee that information will flow along known paths. However, they are inflexible if the environment or team composition changes.

In a dynamic communication structure, the connectivity between agents can change – agents can form or break links as needed, or move in and out of each other’s range if these are physical agents. This is common in mobile robotic swarms or ad hoc networks, where at any given time the communication graph might be different. Dynamic networks are more realistic for many scenarios (e.g., drones fly around and may lose contact with one agent while coming in range of another), and they allow agents to reconfigure their coordination patterns for efficiency or fault tolerance. But they introduce challenges in ensuring the overall system remains connected enough for coordination. A hierarchical MAS with dynamic communication might even allow the hierarchy itself to rewire – for instance, if a leader agent goes down, another agent steps up and all former subordinates now link to the new leader (effectively forming new communication links).

Design patterns: In static setups, the broadcast bus or star topology is a pattern where one central node communicates to all others (typical in centralized MAS). A static tree topology is another, often seen in multi--tier control systems (e.g., each sensor agent sends data to a fixed controller agent). On the dynamic side, a common pattern is neighbor--of--the--moment communication: agents communicate with whoever is nearby or relevant at the time. For example, consensus algorithms have been extended to switching topologies, where at each iteration an agent might average with a different set of neighbors than before, as long as the network is connected over time on average. Results in distributed control theory show that even with dynamically changing graphs (agents moving, links failing, etc.), algorithms like consensus can converge if the graph stays sufficiently connected over time. Another dynamic pattern is the contract broker scenario: an agent needing a task done will seek out a collaborator via a discovery process – this means communication links are formed on the fly between requesters and providers, rather than being pre--wired.

Reconfigurability of groups ties into this as well. \cite{dudek1996} early taxonomy of multi--robot systems included group reconfigurability as a dimension, distinct from just communication topology. Group reconfigurability measures how easily the organizational relationships can change – a high reconfigurability system might allow agents to leave or join groups, roles to shift, and communication links to be rewired during runtime. In HMAS context, this could mean the hierarchical tree is not fixed: perhaps agents can change their parent node if they move geographically, or the hierarchy can deepen/shallow out based on load. For example, in an electricity grid MAS, regions might dynamically form clusters under a substation agent if a certain condition arises, then dissolve later. Or in military operations, units might be reassigned under different command as the formation changes.

The communication structure axis thus spans from rigid and predefined to fluid and adaptive. There are trade--offs: a static structure is simpler to secure (you know exactly which channels exist and can harden them) and often simpler to optimize for (since the communication pattern is known, one can schedule messages or bandwidth accordingly). A dynamic structure is more fault--tolerant and can better accommodate heterogeneity (e.g., if one link is weak or an agent is overloaded, the system can route around it by changing connections). In very large systems, dynamic networks can localize interactions to reduce global traffic – agents talk only to relevant peers rather than broadcasting to all via fixed links.

In designing an HMAS, one should consider: do agents communicate strictly within their hierarchical paths (parent--child or sibling--sibling as authorized), or can they form arbitrary connections when needed? Many multi--agent middleware frameworks (like JADE \cite{bellifemine2001fipa}, SPADE \cite{palanca2020spade3}, etc.) allow a mix: they might provide directory services so any agent can find and message any other (dynamic links), but designers often impose a logical hierarchy on top of that (for organizational clarity).

It’s worth noting that information flow (Section 2.2) and communication structure are related but not identical. Information flow was about the direction of information (who sends to whom conceptually), whereas communication structure is about the topology constraints. You could have a static star topology that nonetheless carries both top--down commands and bottom--up reports (so both flows on a static structure). Or you could have a dynamic peer--to--peer network but where the majority of info is flowing up. Thus, these are treated as separate axes.

In summary, this axis raises the question: is the multi--agent hierarchy built on a fixed wiring diagram or can the agents rewire their communication links during operation? We will see in Section 4 how certain domains favor one or the other. For instance, an operations center for drilling might have a mostly static comm structure (specific agents assigned to specific equipment streams), while a drone swarm performing exploration might constantly change who is relaying information through whom (dynamic mesh network). The robustness of hierarchical coordination in dynamic networks is an active area of research – ensuring that even if the hierarchy “graph” changes, the agents can re--establish a shared understanding of who is leading and how to coordinate is non--trivial. Some recent works (e.g., on consensus and formation control with switching graphs) provide mathematical guarantees for basic coordination under these conditions. Designing higher--level protocols that adapt the hierarchy as agents enter/exit or links fail is still a challenge, tying into our discussion of future directions in Section 5.

\section{Coordination Mechanisms and Design Patterns in HMAS}

Having defined the axes of the taxonomy, the next step is to examine how various coordination mechanisms align with these dimensions. Coordination mechanisms are the algorithms or protocols that agents use to manage their dependencies and jointly perform tasks. In a hierarchical MAS, the choice of coordination mechanism often dictates (and is dictated by) the structure along the taxonomy axes. We highlight several key coordination paradigms and discuss how they manifest in hierarchical contexts, citing relevant academic work that has shaped these ideas.

This section focuses on six representative mechanisms in HMAS: Contract Net Protocol (CNP), auction-based allocation, consensus-based agreement, feudal/hierarchical MARL, organization-centric design (e.g., GAIA/AGR), and agent platforms (e.g., JADE/SPADE). Table~\ref{tab:taxonomy_crosswalk} positions each mechanism along the five axes defined earlier.

\FloatBarrier % prevents earlier floats from slipping up into this paragraph

\begin{table}[!htbp] % h=here, t=top, b=bottom, p=float page, !=override
\centering
\small
\caption{HMAS taxonomy cross-walk: mechanisms × axes.}
\label{tab:taxonomy_crosswalk}
\begin{tabularx}{\linewidth}{
  >{\RaggedRight\arraybackslash}p{2.9cm}  % fixed width for mechanism names
  Y Y Y Y Y}
\toprule
\makecell[l]{Mechanism} &
\makecell[l]{Control\\Hierarchy} &
\makecell[l]{Information\\Flow} &
\makecell[l]{Role /\\Delegation} &
\makecell[l]{Temporal\\Hierarchy} &
\makecell[l]{Communication\\Structure} \\
\midrule
Contract Net (CNP) &
Centralized manager allocates tasks per call &
Announcements top–down; bids bottom–up &
Manager / contractor roles per task (manager may rotate) &
Episodic rounds at task granularity &
Broadcast call; unicast awards \\
\addlinespace
Auctions &
Auctioneer centralizes allocation; market-mediated &
Bids/asks flow to auctioneer; awards flow down &
Auctioneer / bidder roles; peer competition among bidders &
Round-based or continuous clearing &
Star around auctioneer (or tiered in hierarchical auctions) \\
\addlinespace
Consensus / Agreement &
No central controller (leader–follower variants exist) &
Local state exchange among neighbors &
Symmetric peer roles (leader optional) &
Iterative updates over time steps &
Graph / mesh among neighbors \\
\addlinespace
Feudal / Hierarchical MARL &
Manager → submanagers → workers &
Goals/specs downward; rewards/summaries upward &
Hierarchical policies / skills across layers &
Options / skills at different time scales &
Tree (manager → coordinator → worker) \\
\addlinespace
Organization (GAIA/AGR) &
Org-defined constraints and authority &
Role protocols define messages and routing &
Explicit roles, groups, organizational relationships &
Policy / timetable driven (e.g., escalation intervals) &
Role-driven channels; org-specified topologies \\
\addlinespace
Platforms (JADE/SPADE) &
Application-defined (framework runtime) &
FIPA-ACL (JADE) / XMPP (SPADE) &
Behaviours / roles per framework abstraction &
Depends on configured behaviours / tasks &
Star or mesh per design; ACL/XMPP broker / topologies \\
\bottomrule
\end{tabularx}
\end{table}

\FloatBarrier % keep subsequent text from sliding above the table

\subsection{Task Allocation and Negotiation Protocols}

One fundamental coordination problem is how agents allocate tasks or resources among themselves. Negotiation protocols provide structured methods for this, and many can be naturally implemented in a hierarchy.

\begin{itemize}
	\item Contract Net Protocol (CNP): Among task allocation protocols, the CNP is the canonical example, already introduced in Section \ref{subsec:control hierarchy} as a centralized pattern. In practice, CNP has been extended for multi-round negotiation, distributed sensing, and manufacturing applications, illustrating how simple broadcast–bid–award cycles can scale to diverse domains.
	\item Auctions and Market--based Coordination: Auctions generalize the task allocation to potentially multiple tasks and resources, using bids reflecting utility or cost. In a hierarchical setting, one can have auctioneer agents at higher layers that allocate tasks among lower--level agents (similar to CNP but possibly more complex bidding). For instance, a power grid MAS might have an auctioneer agent for energy procurement which gathers offers from generator agents. Auctions assume a communication pattern where bids flow up to the auctioneer (bottom--up info) and allocation results flow down (top--down assignment). This often corresponds to a centralized coordination with possibly peer competition among the bidders. Market mechanisms can be quite flexible in dynamic environments – e.g., if a new agent joins, it can start bidding immediately, which naturally integrates it into the coordination without needing reconfiguration of fixed roles.
	\item Consensus and Agreement: When the goal is to reach a common decision (say, agree on a common plan or value), consensus algorithms are key. In hierarchical systems, consensus might be used among peer leaders or across a network of agents to synchronize state. Hierarchies can expedite consensus by reducing network diameter (a leader can rapidly disseminate a tentative decision), but hierarchies can also bias consensus (if the leader’s information dominates). Research on consensus with hierarchical leaders (like leader--follower consensus) shows that having a leader agent can drive the group to consensus on its value, even under switching topologies. However, a strong leader reduces the benefit of consensus (since it effectively becomes centralized). More democratic consensus fits decentralized, peer--to--peer patterns. Notably, consensus has been extended to scenarios with dynamic communication (agents coming in/out) and still achieves agreement if the network remains connected over time. This makes consensus a robust choice for swarms without fixed structure, although its convergence can be slow.
	\item  Partial Global Planning (PGP): A classic coordination mechanism from the 1990s, Partial Global Planning \citep{durfee1991} was designed for distributed AI systems to coordinate via merging local plans into a “partial global plan.” PGP inherently assumed a hierarchical reasoning: each agent plans locally, then communicates with others to adjust and form a more global plan. There could be a coordination facilitator agent that collects these partial plans (centralizing the info flow) or the agents might peer--to--peer share and negotiate adjustments. PGP highlighted how even without an explicit fixed leader, a coordination structure emerges as agents share plan information and adopt roles in subplans. In taxonomy terms, PGP--style coordination often leads to emergent roles (one agent might naturally become a subplan leader if it has the most constraints, etc.) and uses a mix of bottom--up (sharing local plans up to a potential group plan) and peer communication.
	\item Teamwork Models (Shared Intent): Frameworks like the Joint Intentions theory or the STEAM model provide prescriptions for how agents in a team maintain a shared understanding of who is doing what, and help each other. These frameworks often overlay on a hierarchy by assuming a team leader (or each subteam has a leader) who monitors progress and triggers reorganization if needed. In such coordination, agents have commitments to team goals and to inform each other of certain events (e.g., if an agent cannot do its task it tells the team). A hierarchical team might have commitments structured such that lower agents commit to their superior to achieve subgoals, and the superior commits to the overall mission goal. This implies a fixed role with dynamic task context. Implementations of teamwork models (e.g., in RoboCup rescue or military simulations) typically require robust communication of status (bottom--up) and new directives (top--down) whenever something changes, highlighting the need for reliable communication links (thus often assuming a mostly static communication structure within the team).
\end{itemize}

In summary, task allocation and negotiation mechanisms tend to either assume a central negotiation agent (auctioneer, manager) or operate in a more distributed fashion (consensus, peer negotiation). If an HMAS is towards the centralized end on the Control axis, using protocols like CNP or auctions with designated auctioneers makes sense. If it’s towards decentralized, mechanisms like distributed constraint optimization or consensus--based task assignment (where agents iteratively agree on who will do what) are more appropriate. The taxonomy helps here: e.g., if Communication is dynamic and roles are emergent, an auction that assumes a fixed auctioneer might fail, whereas a peer--to--peer task allocation (like each task “finds” its agent via spreading bids through the network) might succeed. Research continues on hybrid approaches too – e.g., hierarchical auctions (one layer of auctions yields sub--tasks that are further auctioned among sub--teams, etc.), combining scalability of hierarchy with efficiency of markets. More advanced negotiation frameworks are emerging leveraging integration with large language models. An example of this is The AutoGen framework by \cite{wu2023} which is a multi-agent conversation through natural language, which could revolutionize how task allocation is negotiated and conducted.

\subsection{Coordination Through Organizational Structures}

Another perspective is to leverage the organization itself as a coordination mechanism. That is, instead of ad--hoc negotiation, agents coordinate by following rules and patterns defined by their organizational roles and relationships. This aligns with the concept of organizational design in MAS – by pre--defining how agents should interact (who supervises whom, who shares information with whom), the system can coordinate more implicitly.

\begin{itemize}
	\item Hierarchies and Holarchies: In a strict hierarchy, coordination is often achieved by command and control – higher agents assign tasks, lower agents report status. This simple scheme can solve many coordination needs (eliminating conflicts by fiat of the leader, synchronizing actions by scheduling, etc.). Holarchies (hierarchies of wholes and parts, popular in manufacturing systems) allow a group of agents to form a higher--level agent (the “holon”). Coordination happens inside a holon via some protocol, but to outside entities the holon acts as one unit. This is a way of scaling coordination – cluster agents into holons so that inter--holon coordination is simpler (fewer entities). JADE’s organizational abstractions or other agent middleware with groups support this. For instance, a holonic manufacturing system might have a holon for each workshop (with machine agents inside), and a plant--level agent coordinates the holons. This is inherently hierarchical and largely static in structure, leaning on the organizational boundaries to limit coordination scope.
	\item Team and Coalition formation: Sometimes agents spontaneously form a team to handle a task (a coalition). Organizational coordination would say: once the coalition is formed, assign a team leader, define roles, etc., effectively instantiating a temporary hierarchy. The process of coalition formation can be seen as a coordination mechanism to break a complex problem into manageable sub--problems tackled by each coalition. Researchers like  \cite{Horling2004multiple} surveyed paradigms including teams, congregations, societies. Each carries implicit coordination mechanisms. For example, team--oriented plans (TOP) rely on team members adopting roles in the plan; as long as they follow the plan and the communication expectations (like announcing when done with a task), the team stays coordinated. This requires the roles to be defined (possibly at runtime). A hierarchical team (with a captain agent) might coordinate by the captain monitoring team progress and reallocating tasks as needed – essentially an internal CNP within the team.
	\item Norms and Protocols: In more open multi--agent systems (like agents from different organizations interacting), coordination might be achieved via social norms or standard protocols. For example, in a power grid market, all agent participants might follow a standard bidding protocol at given times – this is a temporal coordination (everyone synchronizes on market rounds) and a procedural one (everyone abides by the rules). Hierarchies come into play if, say, there’s a regulator agent that enforces norms or a hierarchy of compliance (smaller agents report to aggregator agents who ensure they all followed the norm). Normative MAS often encode obligations and permissions for roles, which is effectively designing coordination into the system by governing agent interactions. If each role in a hierarchy knows its norms (like “a regional controller must defer to the national controller’s load shed commands”), the agents can coordinate without further negotiation because the norm prescribes what to do.
	\item  Shared Representations / Blackboard: Mentioned briefly under info flow, a blackboard system is where agents coordinate by posting and reacting to a shared data structure (often managed by a central blackboard controller). This is a form of indirect coordination or stigmergy (to draw an analogy from social insects leaving pheromone traces). In hierarchical terms, you might have local blackboards for sub--teams and a global blackboard for the whole system. Agents coordinating via a blackboard follow certain patterns (e.g., when a subtask is solved, write it on the board, other agents will pick up the next needed subtask). The hierarchy here might be in terms of scopes of blackboards (like each level has its own board) which reduces interference – lower--level details stay in local boards, only summaries go to the global board. The design of such systems is often knowledge--intensive but can yield robust coordination as agents adapt to the state of the blackboard continuously.
\end{itemize}

The above mechanisms show that sometimes coordination is baked into the structure and rules of the MAS, rather than requiring heavy runtime negotiation. A well--designed organization – with appropriate communication links, role definitions, and protocols – can make coordination emerge naturally from each agent following its local rules. For example, if every low--level agent knows to execute commands from its superior and to report any anomalies up, and every superior knows to re--plan if a report indicates an issue, the whole organization can respond to a problem with minimal extra messaging. \cite{Horling2004multiple} noted that different organizational paradigms come with inherent coordination patterns (e.g., markets coordinate via price signals, federations via contract and authority, etc.).

In applying the taxonomy, using organizational coordination usually implies a relatively fixed role and communication structure (since the organization is defined). It might also imply certain info flows (e.g., in a market, mostly bottom--up price offers and top--down allocations). However, sophisticated organizational MAS can allow dynamic restructuring (moving towards emergent roles if needed). One research example is Organizational Self--Design, where agents actively rearrange their hierarchy or roles to improve performance (e.g., forming new subgroups or electing new leaders when workload changes). This blurs into the emergent category but guided by meta--level norms on how reorganization can occur.

\subsection{Example Mapping of Mechanisms to Taxonomy}

To concretize the interplay of taxonomy axes and coordination mechanisms, consider a few examples and map them:

\begin{itemize}
	\item Swarm Flocking (\cite{reynolds1987} Rules): This is decentralized (no central control), peer--to--peer info flow (each agent looks at neighbors), no explicit roles (homogeneous, emergent leaders if any by situation), a single time scale (all continuously update positions), dynamic comm (neighbors change as agents move). Coordination mechanism: implicit via local rules (align, avoid, cohere). Taxonomy: fully decentralized, peer info, emergent roles (none fixed), no temporal layering, dynamic network. The swarm achieves coordination (flock movement) without a hierarchy – demonstrating one extreme.
	\item Firefighting Team MAS: Imagine a system where a chief agent allocates fire zones to firefighter agent teams. The chief is central (control), sends assignments (top--down info), firefighters report status (bottom--up). Roles are fixed (chief vs firefighter). Temporal: chief plans overall strategy (longer horizon), firefighters make split--second decisions in their zone (short horizon) – a temporal hierarchy. Communication: mostly static (pre--assigned teams, radio links). Coordination mechanism: task allocation by chief (like CNP or instruction), plus teamwork within each team (maybe a squad leader among firefighters). Taxonomy: centralized/hybrid control (chief at top), info mainly top--down \& bottom--up, roles mostly fixed, temporal layered, static comm. This system coordinates through structured command: efficient but if chief node fails, could be problematic (unless there’s succession).
	\item Smart Grid Energy Management: A modern example: a distributed energy grid with prosumers. Suppose each neighborhood has an agent that manages local resources (solar panels, batteries) and there’s a city--level agent that balances load among neighborhoods. This is a two--level hierarchy (hybrid control: neighborhood agents make many decisions locally, city agent intervenes for global balance). Info flow: neighborhoods report forecasts up, city sends back demand limits or prices (bottom--up sensing, top--down directives). Roles: fixed (neighborhood vs city roles), but perhaps emergent if neighborhoods can self--organize trading. Temporal: city agent might plan day--ahead (long horizon), neighborhood agents adjust in real--time (short horizon) – clear temporal stratification \citep{dragomir2025}. Communication: could be dynamic if the grid reconfigures (e.g., microgrids islanding and reconnecting), but generally the structure (neighborhood to city) is fixed. Coordination mechanisms could include iterative price setting (city agent sets a price signal, neighborhood agents respond with demand adjustments, repeated until balance – a kind of distributed negotiation with the city agent as mediator). This maps to mostly hierarchical but using a market--like mechanism internally. Taxonomy: hybrid control, mixed info flow, fixed layered roles, temporal hierarchy, mostly static links (neighborhood to city). Real--world experiments have shown such hierarchical MAS can improve grid resilience and scalability, by reducing single points of failure and integrating distributed resources efficiently .
\end{itemize}

These examples illustrate how one can use the taxonomy axes to reason about the design choices and appropriate coordination methods. In general, aligning the coordination mechanism with the structural design is crucial: a mismatch (e.g., trying to do fully decentralized consensus in a system that is architected with a strict central hierarchy) can lead to poor performance or even system contradictions. Conversely, understanding that an architecture can enable certain types of coordination can inspire using those mechanisms (e.g., if you have a powerful central agent, leverage it to run global optimization or scheduling). The next section will delve into concrete application domains, which will further show how these patterns manifest in practice.

\section{Industrial Applications of Hierarchical MAS}
\label{sec:HMAS in Industry}

A major motivation for studying HMAS is their applicability to real--world complex systems. This section explores several domains – with a focus on the energy sector and operations management – where hierarchical multi--agent frameworks have been applied or could be beneficial. For each domain, we describe the problem context, how a hierarchical MAS approach has been or could be structured, and any reported results or challenges from the literature. We pay particular attention to the oil and gas industry as requested, providing depth on how HMAS can optimize drilling and production operations. We also touch on smart grids (energy distribution), logistics and manufacturing, and human--in--the--loop operations.

\subsection{Smart Grids and Energy Management}

The transition to smart grids and distributed energy resources has created a complex control problem well--suited to multi--agent solutions. Energy systems naturally have hierarchical structure: for example, devices (smart appliances, EV chargers, solar panels) connect to local controllers (home or building energy management systems), which connect to feeder or microgrid controllers, then up to regional and national grid operators. Researchers have leveraged this by deploying MAS at multiple levels of the grid control hierarchy . Each agent can make decisions appropriate to its scope – e.g., a home agent optimizes appliance usage to minimize cost, a neighborhood agent coordinates power flow among homes to maintain voltage stability, and a central agent ensures global supply--demand balance .

One example is a decentralized hierarchical MAS for sustainable energy management proposed by \cite{dragomir2025}. They implemented a three--layer agent system: device--level agents, a microgrid (neighborhood) agent, and a main grid agent . The system performed real--time monitoring and control of a simulated power grid with renewable energy penetration. By using a hierarchical architecture, they could balance multiple objectives (cost, environmental impact) effectively, and the system proved more resilient and scalable than a purely centralized control . Notably, the hierarchical MAS reduced single points of failure and allowed easier integration of new resources, since each microgrid agent could manage its domain autonomously and only critical information was sent to the top .

Coordination mechanisms in such energy MAS often mix market--based and cooperative strategies. For instance, multi--agent energy trading platforms treat each prosumer agent as an economic actor that can buy/sell energy; a higher--level broker agent or auction mechanism sets prices that coordinate the agents. Alternatively, load balancing can be done cooperatively: neighborhood agents might collectively maintain grid frequency by slight adjustments in load or battery usage, coordinated through a leader or consensus on frequency deviation. A survey by \cite{moradi2016} provided a taxonomy of MAS applications in power engineering, noting use cases like fault diagnosis, voltage control, market trading, and load restoration. Many of these solutions inherently use hierarchical concepts – e.g., protection systems have local relays (agents) and central coordinators, restoration systems have field crews guided by control center agents.

A specific case of HMAS in energy is repair crew dispatch for power networks. \citep{Qui2023multiple} formulated a hierarchical multi--agent reinforcement learning for dispatching repair crews in a coupled power--gas grid after failures. They treated high--level scheduling and route planning separate from low--level crew movement decisions, training agents in a two--level MARL framework. This separation (temporal and role hierarchy) helped tackle the large state--action space and yielded efficient restoration strategies for the network. It highlights how hierarchical control (here, planning vs execution layers) can reduce complexity in multi--agent optimization problems like grid maintenance.

Real--world deployment of HMAS in grids is still in early stages, partly due to safety and reliability concerns. The power industry typically requires extensive validation; thus many MAS controllers are tested in simulations or pilot projects. A challenge noted by several studies is the integration of MAS with legacy grid control systems. Agents need to work with existing SCADA and human operators, requiring transparency. This calls for HMAS designs that produce explainable actions (so operators trust their decisions) and adhere to grid operating norms (e.g., emergency cutoff rules). Hierarchical arrangements help here: a human operator can interface with the top--level agent (rather than hundreds of device agents), making oversight more manageable. In essence, the MAS hierarchy can be aligned with the human supervisory hierarchy, a theme revisited under human--agent collaboration.

\subsection{Oil and Gas Operations (Autonomous Drilling and Production)}

The oil and gas (O\&G) industry is characterized by large--scale operations, distributed assets (wells, pipelines, facilities), and the need for real--time decision--making for safety and efficiency. Historically, O\&G has been slower than some industries in adopting multi--agent autonomy. However, the concept of the “Digital Oil Field” or intelligent oilfield has gained traction, where AI agents continuously monitor and optimize operations. Hierarchical MAS can play a crucial role in such settings by mirroring the structure of operations centers and field units.

Consider an oil field with multiple wells and a central field operations center. One could design an HMAS where each well (or each drilling rig) has an autonomous agent managing that asset – tuning parameters, detecting anomalies, controlling equipment – and higher--level agents at the field or asset level coordinate between wells (e.g., to optimize overall production or manage shared resources like pumps or power). Onshore field management was explored by \cite{oliveira2013} in their SGCP system. In SGCP, each oilfield component is modeled as an agent: operational agents for production equipment, intervention agents for maintenance, and supply agents for inventory/logistics. They form a multi--layer network that cooperatively diagnoses production problems, analyzes profitability, and supports decision--making by field engineers. For instance, an agent monitoring a particular well’s flow might detect a drop in output (using statistical control charts) and send an alert requesting an intervention (like a maintenance check). The intervention agent would coordinate scheduling a crew or adjusting a choke valve. Meanwhile, a higher--level agent could assess if rerouting flows or adjusting other wells can compensate for the drop. This system essentially encapsulated expertise of different functions into different agent types and let them collaborate through an internal economy of “adding value” and “transferring costs”. While the deployment details were not widely reported beyond prototypes, it demonstrates the viability of a hierarchical decomposition: site--level agents manage local tasks, and a top--level agent (or simply the emergent result of the network) handles holistic field optimization.

In drilling operations, which are highly complex and dangerous, multi--agent approaches have been proposed for drilling automation. A drilling rig can be seen as a multi--agent environment: one agent controlling the drill bit, another managing mud flow, another monitoring vibrations, etc., all under a supervisory agent that ensures the overall drilling process stays within safety and performance bounds. O\&G companies like Shell have invested in “real--time operations centers” where human experts remotely assist multiple rigs. This can be augmented with agents: each rig sends data to a rig--agent at the operations center, which uses machine learning to detect warning signs (like kick or stuck pipe indicators). That rig--agent might then either alert a human or directly command the rig’s local agents to take corrective action (e.g., adjust drilling parameters). Hierarchy is clear: rig--site agents handling second--by--second control, overseen by a central agent (or team of agents) handling minutes--to--hours scale decisions and coordination among rigs (like prioritizing resource allocation or crew movements if multiple rigs need attention).

One interesting application is preventive maintenance and asset management. Oilfield assets often degrade or fail, and predicting these events can save millions. A hierarchical MAS could integrate: low--level monitoring agents on equipment (pumps, compressors) doing anomaly detection, higher--level agents scheduling maintenance tasks when multiple alerts come in (optimizing which equipment to fix first given limited crew), and top--level agents balancing production targets with maintenance downtime. A multi--agent approach can naturally handle the distributed nature of assets and the need for local autonomy (equipment should shut down if it’s about to explode, without waiting for central approval, but scheduling its restart or replacement can be done centrally).

Despite successful prototypes, adoption in O\&G has been slow. \cite{hanga2019} surveyed AI and MAS in O\&G and noted that while MAS showed good performance in simulations (for supply chain optimization, facility management, etc.), they “have not gained the expected popularity among oil and gas companies yet”. They suggest reasons such as conservatism in the industry, the complexity of integrating with existing systems, and the need for more proof of reliability. They argue that embedding advanced ML (machine learning) into MAS could unlock more value (for example, using ML to predict well behaviors and MAS to coordinate responses). This is an active area of research: how to build hybrid ML--agent systems that can learn from the vast data in oilfields and still adhere to the operational constraints. One approach is to use ML to develop predictive models (say of equipment failure or of reservoir response) and use multi--agent decision systems to act on those predictions hierarchically.

A concrete futuristic scenario: imagine autonomous drilling rigs coordinated by an intelligent operations center. Downhole sensors and surface equipment are managed by agents on the rig (with maybe a drill bit agent, a mud system agent, etc.). They handle second--by--second control using reinforcement learning policies that maximize rate of penetration while avoiding dangerous differentials. At the operations center, a supervisory agent monitors all rigs in the field, reallocating shared resources like power or alerting rigs of upcoming changes (e.g., instructing one rig to pause if another hits a critical stage requiring extra monitoring manpower). Human managers oversee the top--level agent, who provides them high--level summaries and explanations (“Rig 7 slowed drilling due to high vibration, Agent adjusted weight on bit by 5\% to mitigate – outcome successful”). This hierarchical autonomy could dramatically increase efficiency (agents react faster than humans to transient events) and safety (constant monitoring with automated shutdown triggers). The taxonomy axes for this would be: control is hybrid (autonomy on rig, with human--in--loop oversight), info flow bottom--up (data to center) and top--down (commands to rig), roles semi--fixed (rig agent vs center agent vs human roles defined, but within rig agents could dynamically take lead in different sub--tasks), temporal layered (real--time control vs strategic field scheduling), communication partly dynamic (e.g., rigs might collaborate or share info peer--to--peer in some cases like exchanging learnings about rock strata – although likely via center).

A challenge in O\&G MAS is the harsh environment and reliability. Communication links can be slow or disrupted (especially offshore or remote fields), so agents must handle degraded communication – a strong case for giving local agents significant autonomy (they can act even if cut off from central). It also demands robust error handling and fail--safes; agents controlling physical processes must adhere to strict safety rules. This ties to trust: convincing engineers to trust agent decisions is non--trivial. Transparent logic (perhaps rule--based agents or at least explainable ML) and extensive testing are needed.

In summary, hierarchical MAS in oil and gas operations show promise for integrated asset management, by aligning with the sector’s multi--level operational structure. Some prototypes like SGCP have demonstrated feasibility in onshore fields. The full potential – e.g., completely autonomous drilling optimization – is still being researched. The industry’s cautious approach means HMAS solutions must emphasize reliability, explainability, and ROI to gain traction. As digital transformation progresses (with companies like Shell, BP digitizing assets), we expect more trials of MAS coordinating, for example, fleets of autonomous inspection robots or optimizing entire well lifecycle (from drilling to production to maintenance) through an agent--based orchestration. The taxonomy provides a conceptual map for these developments, ensuring that designers cover the needed aspects (communication reliability, role definitions, etc.) when building O\&G agent systems.

\subsection{Warehouse Automation and Logistics}

In the logistics domain – encompassing warehouses, distribution centers, and transportation networks – multi--agent coordination has already made significant inroads. Companies like Amazon and Alibaba operate automated warehouses with swarms of robots (AGVs – Automated Guided Vehicles or AMRs – Autonomous Mobile Robots) for moving goods. Typically, a hierarchical control is present: a central management system assigns tasks (like “Robot X, fetch item Y from location Z”), and the robots handle the navigation and local avoidance. However, as these systems scale, more autonomy is being given to the robots collectively to self--coordinate and optimize throughput.

A recent survey noted MAS are widely used in warehouse automation, citing examples like Amazon Robotics and Cainiao (Alibaba’s logistics). In these scenarios, autonomous agents such as robots, conveyors, and shuttles collaborate to streamline tasks like picking, sorting, and transporting goods. Coordination is essential to avoid conflicts (collisions, congestion) and to optimize efficiency (preventing multiple robots from going to the same shelf unnecessarily, etc.). Usually, a multi--tier MAS is deployed: for instance, section controllers manage a zone of the warehouse, while each robot is an agent within that zone. The section controller can reroute robots if it detects a jam and can balance workload by holding some robots idle if too many are in one area. At the top, a warehouse management agent (or system) might assign orders to sections, effectively playing an orchestrator role.

One interesting coordination mechanism in warehouses is the use of priority--based schemes and negotiation among robots. For example, when two robots approach an intersection, rather than a fixed traffic light, they could negotiate right--of--way: possibly one robot acts as a local “intersection manager” agent dynamically (like a token leader role). Alternatively, robots could use simple priority rules (encoded as norms: e.g., robot with heavier load or higher urgency gets priority). These rules are part of the MAS design that ensure safety and deadlock avoidance. The hierarchical angle comes when these rules can be overridden by higher--level logic in exceptional cases (say a supervisor agent sees that a high--priority shipment is at risk of delay, it might instruct all robots to yield to the one carrying that shipment, essentially injecting a top--down priority override).

Drone delivery is another logistics area benefiting from hierarchical MAS. \cite{dorling2017}  and others studied vehicle routing for delivery drones, which can be viewed as a multi--agent path planning problem with constraints \citep{Horling2004multiple}. A central system could plan routes for all drones (like solving a big optimization), but a hierarchical approach might break it down: a central agent allocates delivery tasks to drones (task allocation), then drones plan their own routes or coordinate with nearby drones to avoid collisions or share air corridors. \cite{Frachtenberg2019} discussed scaling drone delivery and likely touched on decentralized coordination to avoid air traffic issues. In an urban environment, one could envision area controllers (like air traffic control sectors) as agents that manage local swarms of drones – a hierarchical control borrowed from manned aviation.

Transportation systems (fleet management, traffic routing) similarly use hierarchical control. For example, city traffic MAS might have intersection agents controlling traffic lights and a central traffic agent optimizing citywide flow or responding to incidents. Self--driving cars can be seen as agents that might one day coordinate with each other without centralized control; however, practical considerations suggest a hybrid approach where city or road infrastructure provides guidance (like variable speed limits, or coordination messages) to the autonomous vehicles which then locally adjust. Sun et al. (2025) survey notes such mixed approaches for coordinated autonomous driving, including selecting regional leader vehicles to optimize regional traffic as a hierarchical strategy.

From the taxonomy perspective, most warehouse/logistics MAS are hybrid: they use centralized algorithms for high--level optimization (e.g., order batching, overall routing plans) and decentralized agent coordination for execution (robots negotiating paths, trucks timing their routes to avoid bunching). Communication is often a mix: a central server communicates with all robots (static star for assignments), but robots also share peer--to--peer signals for collision avoidance (dynamic local networks). Roles might be largely fixed (robot vs supervisor roles), though emergent roles appear if, say, one robot temporarily takes on a leader role in a platoon of delivery vehicles.

The success in this domain is evident: Amazon’s fulfillment centers reportedly handle hundreds of thousands of orders using robot swarms coordinated by multi--agent scheduling algorithms, achieving efficiency beyond human--only operations. Key performance gains come from reduced idle time and congestion as agents communicate their positions, share task updates, negotiate resource usage, prevent bottlenecks, and minimize idle time by seamless teamwork. Essentially, MAS coordination ensures the warehouse operates as a harmonious whole rather than isolated robots.

A challenge going forward is integrating human workers with these agent swarms, as many warehouses still have humans doing certain tasks. Hierarchical MAS could include human--agent teams (for instance, a human picker working with a robot that brings shelves – the robot agent and human form a team coordinated by a higher--level system that assigns them complementary tasks). Ensuring safety and clear communication in these human--agent interactions is paramount – the MAS must be designed to avoid accidents (e.g., robot gives right--of--way to human always) and to be transparent (human knows what the robot will do next, perhaps via signals or AR interfaces).

\subsection{Human--Agent Collaboration in Operations Centers}

In complex operations whether emergency management, military command, or industrial control rooms – it is common to find human experts and AI agents working side by side. Hierarchical MAS can facilitate this by positioning human operators at certain decision nodes in the hierarchy and automating others. Essentially, the human becomes an agent in the MAS (with unique capabilities like judgment, ethical reasoning, creativity) and the system must be designed for effective coordination between humans and agents.

A common approach is to give humans supervisory roles over agents (which parallels fixed roles in hierarchy – e.g., human is always the ultimate decision--maker on critical actions). For example, in a real--time operations center for drilling (as mentioned in 4.2), a human manager might oversee 5 rig agents. The HMAS would require the rig agents to get approval from the human for any action beyond certain thresholds (like shutting down a well). In normal conditions, the human trusts the agents to handle routine tasks, but the agents continuously explain their status and intentions to the human. This requires a high level of explainability; research on explainable agents is looking into how to produce human--friendly summaries of agent decision logic, especially when using opaque models. Some approaches use layered communication for explainability: e.g., an agent might first issue a simple alert (“Well \#3 pressure rising, taking action to reduce flow”) and if the human inquires further, the agent can provide a more detailed rationale or data plot. The hierarchical setup can actually help manage explainability, because the top--level agent can collate and filter information from lower--level agents, presenting only what’s relevant to the human (preventing information overload in complex MAS).

Human--swarm interaction (HSI) is a subfield studying how a human can effectively command or cooperate with a swarm of agents (like drones or robots). One technique is to allow the human to issue high--level commands to a lead agent or a subgroup, effectively treating the swarm hierarchically so the human isn’t micromanaging every unit. For instance, a human supervisor might tell a drone swarm agent “survey this area for 10 more minutes then return,” and that agent disseminates the plan to the others. Alternatively, the swarm could self--organize but the human can inject some influence, such as attracting the swarm’s attention to a region by clicking on a map (some swarm agents take on the role of propagating that influence to others – a bit like emergent hierarchy where the human’s input is treated as coming from a virtual leader).

A powerful benefit of HMAS for human collaboration is handling heterogeneity. As Sun et al. note, heterogeneous MAS – including humans as a special type of agent – bring flexibility, but coordination is harder because not all agents follow the same protocols or capabilities. A hierarchical framework can accommodate heterogeneity by grouping similar agents under specialized mediators. For example, human team members might be managed by a human liaison agent that communicates in natural language and interprets the human’s intent to the rest of the MAS, whereas robotic agents might be managed by a different agent that handles their telemetry. This way, each type of agent (human or machine) “talks” to a suitable counterpart, and those counterparts coordinate. This concept was explored in some human--robot team simulations where a proxy agent represents the human in the agent society, making it easier for the autonomy to accommodate human input which can be high--level or unstructured.

Trust is a critical issue in human--agent teams. Humans need to trust agents to delegate autonomy, and agents (if sophisticated enough) need to trust human directives (especially if they have some learning or adaptive capacity that could reject harmful commands). Computational trust models have been investigated to adjust agent behavior based on trustworthiness of information sources . For example, if a human operator is known to often override the agent’s decisions incorrectly, the system might adapt by seeking additional confirmation or by providing more evidence to persuade the human. Conversely, if an agent has performed poorly recently, the human will lose trust unless the system can explain and improve. As \cite{pinyol2013} and others showed, trust--based role assignment can even determine who (human or agent) takes lead in a subtask dynamically. In a hierarchical context, perhaps the human normally approves plans (like a high--level checkpoint), but if trust is high, the agent might execute some plans without bothering the human (adjusting autonomy level).

A concrete illustration is emergency response coordination. Imagine a multi--agent system managing a building fire: sensor agents, drone scouts, firefighting robot agents, and human firefighters all collaborate. A hierarchy might have a central emergency coordinator agent advising the incident commander (a human). Sensor and drone agents feed data up (bottom--up), the coordinator agent and commander analyze and decide high--level strategy (e.g., where to focus resources), then commands go out to robot agents and firefighters (top--down). Human firefighters might each have an assistant agent on their devices that gives suggestions or routes to follow, essentially a human--agent pair per team. If a firefighter doesn’t follow a suggestion, the agent notes that (maybe adjusting its plan or marking that route as blocked – an example of human override the agent must accept). This is very much a human--in--the--loop hierarchical MAS. The team process is facilitated by MAS: one study on emergency management teams indicated that having a multi--team coordination layer improved communication between local teams and state--level coordination. That extra layer can be an agent that accumulates field info and suggests reallocations of responders between incidents.

In all these cases, hierarchical MAS design provides points of interface for humans and agents, rather than a free--for--all. Humans typically can’t or shouldn’t monitor every single agent in a swarm, so hierarchy funnels interactions: e.g., one human supervising one agent who supervises 50 lower agents. This is analogous to military or corporate structures (span of control management). It also aids training and simulation: one can simulate the MAS with or without human inputs to see how it behaves, then insert human agents and see how outcomes change, to identify optimal role boundaries.

Finally, a forward--looking challenge is integrating LLM--based agents (large language model AI, like ChatGPT--based agents) into human--agent teams. These LLM agents are quite adept at communication and could mediate between humans and more specialized robotic agents. For example, an LLM agent could take a human commander’s intent in natural language and translate it into formal goals for the MAS, and conversely summarize MAS data into a briefing in human language. Early work on LLM--based multi--agent systems has started exploring how to align such agents with system objectives. A multi--dimensional taxonomy for LLM--agent architectures suggests classifying how much autonomy vs. human alignment they have in each aspect (goal management, communication, role delegation, etc.). The integration of these powerful reasoning agents could greatly enhance human--agent collaboration by bridging the gap between unstructured human instructions and structured agent actions. However, it also introduces risk if the LLM agent generates incorrect or ambiguous commands – again highlighting the need for oversight and clearly defined authority hierarchies (e.g., perhaps never let an LLM agent directly control a physical process without a validation from a rule--based safety agent or a human).

Key takeaways in human--agent collaborative HMAS: keep humans in appropriate roles (neither overload them with micromanagement nor exclude them from critical decisions), ensure agents are transparent and trustworthy, and use hierarchy to manage complexity of interactions. With advances in interfaces (AR/VR for situational awareness, brain--computer interfaces as some studies even explore ), the line between human and agent roles might further blur, making the hierarchical structure even more crucial to prevent chaos and maintain clarity in joint missions.

\section{Open Challenges and Future Directions}

Hierarchical multi--agent systems, while powerful, introduce unique challenges that merit further research. In this concluding section, several open issues and emerging directions are outlined, informed by the latest literature and the discussions throughout this paper. Addressing these will be key to unlocking the full potential of HMAS in real--world and future AI deployments.

\subsection{Trust, Accountability, and Human--in--the--Loop Integration}

The paper discussed how trust is essential in human--agent hierarchies. Open challenges remain in calibrating trust – ensuring humans neither over--trust (leading to complacency when the agent might be wrong) nor under--trust (leading to ignored agent advice and lost benefits). Computational trust models \citep{pinyol2013} that assess agent confidence and past performance can feed into adaptive autonomy: e.g., an agent could say “I am 90\% confident in this plan; I recommend it, but since it’s critical, please approve” versus “I have done this routine task many times with success, proceeding autonomously.” Some works propose agents explicitly model the human’s trust and adjust behavior (explain more or seek approval more often if trust is low, or quietly handle things if trust is high). Designing these behaviors without causing annoyance or confusion is tricky – the system needs to be predictable in how it involves the human. Establishing accountability is part of trust: in a hierarchy, if something goes wrong, who (or what agent) is responsible? Ensuring each agent’s actions are logged and traceable to directives from above (or exceptions it handled) will help assign responsibility. This is analogous to an organization where each level is accountable to the one above. Techniques from auditing and verification of MAS can be applied, possibly using blockchain or secure logs for traceability of multi--agent decisions.

For human--agent teams, another area is improving the communication interfaces. Natural language interaction is very promising here – with advanced LLMs, agents can brief humans or answer questions in fluent language, which could greatly reduce training needed for operators to use these systems. Recent efforts like Autogen \citep{wu2023}  demonstrate multi--agent conversation programming for LLM applications. This blurs the line between a human giving an order and programming the MAS – a user might simply say, “Make sure all wells are stable before increasing output” and the top--level agent interprets and disseminates that. The taxonomy by \cite{handler2023} on LLM--based multi--agent systems introduced aspects of autonomy vs. alignment across different viewpoints of the architecture, highlighting that not only must LLM agents be autonomous, they must align with human intentions and constraints at each decision point. So a future HMAS might incorporate an LLM agent at the top to mediate between human and machine layers, carefully constrained to not go out of bounds.

\subsection{Scalability and Hybrid Coordination Strategies}

While hierarchy is one solution to scalability, extremely large--scale systems (think thousands or millions of agents, like IoT devices or swarms) push current approaches to their limits. A hierarchy can only reduce complexity to a point; beyond that, new concepts might be needed, such as dynamic hierarchical clustering – automatically restructuring the hierarchy as the scenario evolves. Research in large scale UAV swarm confrontations have demonstrated how hierarchical structures can maintain performance even when scaled, work by \cite{chunyang2025} used fuzzy reinforcement learning to keep the computational and storage requirements manageable. This overlaps with the communication axis: dynamic networks could include dynamic hierarchy formation. Research could focus on algorithms for agents to self--organize into hierarchies optimally. For instance, in a disaster response with hundreds of robots and humans arriving, how do they decide who coordinates whom? Perhaps initially a flat peer--to--peer works until a leader emerges (maybe the one with best situational overview), then that leader organizes units under it. Mechanisms like leadership election, coalition formation, and role rotation become critical in such open MAS. These dynamic hierarchies should ideally maintain most of the benefits of designed ones (e.g., not oscillating roles too often, ensuring clear authority at any time). There is theoretical work on forming spanning trees for communication and such, but incorporating mission goals into the formation of an efficient hierarchy is an open problem.

The hybridization of hierarchical and decentralized coordination has been identified as a promising direction. Future systems might not choose one approach but switch modes depending on context. For example, in nominal conditions, a strict hierarchy could coordinate routine tasks (fast and predictable), but in a novel or emergency situation, the system might relax hierarchy to let agents self--organize and brainstorm solutions (more exploration). Once a solution is found, it might re--impose a hierarchy to execute it. How to detect those conditions and fluidly change coordination modes is an interesting question. This touches on meta--coordination: agents coordinating about how to coordinate. Some initial frameworks allow agents to negotiate organizational changes (e.g., “I propose we elect a leader for this task” or “let’s operate in team mode vs market mode now”). Formalizing and automating such meta--decisions could lead to highly adaptive MAS that sometimes behave like swarms and other times like rigid hierarchies as needed.

Scalability also concerns computation: solving assignments or plans even at one layer can be hard when there are many agents. Here, approximation algorithms and heuristics integrated with learning are likely needed. Hierarchical reinforcement learning itself has scaling issues if not carefully structured (the state space at high levels might still be huge). Combining classical optimization (for high--level coarse allocation) with learning (for low--level fine control) is one promising angle – use the strengths of each at appropriate layers.

\subsection{Integration with Learning Agents and LLMs}

The rise of large language models and other advanced learning agents opens new opportunities and challenges for HMAS. On one hand, these agents can bring greater cognitive abilities – they can reason with knowledge, generate plans, and communicate richly. On the other hand, they are stochastic and not entirely reliable (LLMs can hallucinate facts or be misaligned with exact goals). Integrating them into a safety--critical hierarchical MAS (like controlling physical infrastructure) must be done with caution. A potential architecture is to have LLM--based agents in advisory or planning roles, but have traditional rule--based or model--predictive controllers in execution roles. For example, an LLM agent could suggest a new strategy to optimize an oil production schedule (based on reading historical reports and figuring out patterns), but the actual control to implement that strategy is handled by a proven algorithm that the LLM’s output is fed into. Alternatively, LLMs could be used to dynamically generate coordination protocols or role definitions on the fly in novel scenarios (because they have “seen” many examples in training).

A recent taxonomy by \cite{handler2023} tries to categorize LLM--powered multi--agent architectures by how they balance agent autonomy and system alignment with human or designer intent. This underscores a future direction: ensuring that powerful autonomous agents remain aligned with overall system objectives, especially when those objectives might be hard to encode fully. Hierarchical structures can embed alignment by design (the top--level agent ensures alignment and constraints), but if the top--level agent itself is an LLM or learned model, one might need an alignment layer above it – possibly a human or a symbolic checker. One could imagine a hierarchy where the very top is a rule--based “safety governor” agent (or a human overseer for critical matters), middle layers are learning--based planners (LLMs or neural policies), and bottom layers are hard--coded controllers – a kind of safety envelope around learning. Achieving seamless coordination among such heterogeneous decision paradigms is an open challenge (how does a rule--based agent communicate with an LLM agent effectively? Are they using a common language or ontology? How to prevent the LLM from misinterpreting an oversight rule?).

Multi--agent learning itself is a big topic. Many MARL (multi--agent reinforcement learning) algorithms struggle with non--stationarity and scalability. Hierarchy has been proposed as a solution (e.g., hierarchical MARL, manager--worker as in Feudal networks, role discovery as in ROMA). Future research will likely delve into curriculum learning for hierarchies: training lower layers first, then higher, or vice versa, or alternating – to stabilize learning. Also, transfer learning in HMAS: if you add a new agent or a new layer, can it learn from the behaviors of previous similar ones? For instance, if a company adds a new regional control agent above existing city agents, can it use data from how city agents interacted to quickly set its policy? This is analogous to how human organizations promote someone and expect them to learn from history.

Lastly, ethics and security in hierarchical MAS is an emerging concern. A malicious agent at a high level could do much damage by sending bad commands, so building in robust authentication, redundancy (maybe a voting among high--level agents), and detection of sabotage is crucial, especially as MAS run critical infrastructure. Hierarchy can either mitigate or worsen this – a strict hierarchy might allow a single point of attack (the leader), whereas a decentralized approach has no single failure but is harder to secure globally. Research might explore hybrid secure architectures (e.g., multiple redundant leader agents, or consensus among leaders for critical decisions).

In conclusion, hierarchical multi--agent systems sit at the intersection of classic organizational principles and cutting--edge AI techniques. The coming years will likely see these systems expand in capability – handling more agents, more learning, and more complex tasks – while researchers work to ensure they remain safe, transparent, and aligned with human values and goals. By addressing the challenges outlined above, we can move towards HMAS that not only perform efficiently at scale but also earn the trust of the people who rely on them. The taxonomy and analyses provided in this paper aim to guide both the understanding of current systems and the design of next--generation hierarchical multi--agent architectures that will underpin increasingly autonomous industrial and societal systems.

\section{Conclusion}
In this paper, a comprehensive examination of hierarchical multi--agent systems (HMAS) was presented, proposing a new taxonomy that categorizes these systems along key design axes including control distribution, information flow, role allocation, temporal layering, and network structure. By surveying foundational research and recent developments, the analysis illustrates how these dimensions manifest in various coordination patterns and architectural frameworks. A unifying theme is that hierarchy – in its many forms – offers a powerful means to tame complexity in multi--agent coordination, but it must be wielded with careful consideration of trade--offs: global vs. local control, fixed vs. adaptive organization, long--term vs. short--term optimization, and robust static links vs. flexible dynamic reconfiguration.

The taxonomy is connected to real--world applications in domains such as smart energy grids, oil and gas operations, warehouse logistics, and human--in--the--loop control centers. These case studies underscore that there is no one--size--fits--all hierarchy; rather, effective HMAS design is context--dependent. For example, an onshore oil field benefited from a functional hierarchy of agents that mirrored maintenance, production, and supply departments, whereas a warehouse robotics system relied on a spatially partitioned hierarchy to manage local traffic and global workflow. In each scenario, the taxonomy’s axes helped analyze why those designs succeeded and what limitations remain. Notably, the oil and gas domain, with its entrenched operational structures and safety--critical processes, illustrated both the promise of HMAS (e.g., proactive diagnostics and reduced personnel load) and the challenges (industry adoption lagging due to trust and integration issues). An in-depth look is also provided at how a hierarchical agent approach could revolutionize drilling and production management, drawing parallels to existing digital transformation efforts in that sector.

The exploration of design patterns (like contract nets, auctions, consensus, and organizational rules) in Section 3 further connected abstract taxonomy concepts to concrete coordination mechanisms. The analysis highlights how certain patterns naturally align with certain parts of the design space (for instance, auction protocols flourish under a central broker model with top--down communication, whereas consensus requires peer--to--peer links in a decentralized fashion). This mapping is intended to guide practitioners in choosing the appropriate coordination strategy given the structure of their agent system.

Finally, forward--looking perspectives are addressed in Section 5, identifying open research challenges. Chief among these is the need for greater explainability, trustworthiness, and adaptivity in hierarchical MAS. As these systems become more autonomous and pervasive (e.g., managing national power grids or autonomous vehicle fleets), stakeholders will demand assurances of safety and clarity. Techniques for transparent decision hierarchies, human--aware coordination, and dynamic reorganization will be crucial. The paper also notes the intriguing opportunity to integrate large language models (LLMs) and advanced learning agents into HMAS , which could imbue agent teams with improved reasoning and communication skills – effectively, more human--like collaboration capabilities. Early taxonomy work on LLM--based multi--agent architectures suggests that balancing autonomy with alignment (to goals and norms) will be the key challenge there, and hierarchical control might be one mechanism to impose that alignment.

In conclusion, hierarchical multi--agent systems represent a rich and evolving field at the crossroads of AI, control theory, and organizational science. The taxonomy and analysis provided in this paper aim to serve as a useful framework for both researchers and system designers. By breaking down the design space and citing representative solutions from the literature, the work offers a scaffold to reason about new problems and ensure that novel HMAS architectures learn from past successes and pitfalls. As AI systems continue to scale up and permeate high--stakes industries, the principled design of hierarchies – to maintain order, efficiency, and human compatibility – will likely become even more critical. Hierarchies in MAS are not simply about who commands whom; they encapsulate how information is abstracted, how responsibilities are shared, and how collective intelligence can be orchestrated. Getting those aspects right is essential to creating multi--agent systems that are robust, scalable, and beneficial in the long run.

\bibliographystyle{plainnat}
\bibliography{Bibliography}

\end{document}